\documentclass[referee]{raa}
\usepackage{graphicx,times}
\usepackage{natbib}
\usepackage{amssymb,amsmath}
\usepackage{longtable}
\bibpunct{(}{)}{}{}{}{}

\usepackage[a4paper=true,dvipdfm=true,pagebackref=true]{hyperref}
\hypersetup{pdftitle = The title of my PDF, pdfauthor = My name, pdfsubject= The subject, pdfkeywords = keyword1 keyword2 keyword3}
\hypersetup{colorlinks = true, linkcolor = green, anchorcolor = red, citecolor = blue, filecolor = red, pagecolor = red, urlcolor = red}

\begin{document}

   \title{Photometric observations and light curve solutions of the W UMa stars NSVS~2244206, NSVS~908513, CSS~J004004.7+385531 and VSX~J062624.4+570907}

 \volnopage{ {\bf 2012} Vol.\ {\bf X} No. {\bf XX}, 000--000}
   \setcounter{page}{1}

   \author{D. Kjurkchieva\inst{1}, V. A. Popov\inst{2}, D. Vasileva\inst{1}, and N. Petrov\inst{3}}

   \institute{Department of Physics, Shumen University, 115, Universitetska Str., 1712 Shumen, Bulgaria; {\it d.kyurkchieva@shu-bg.net}\\
 \and
   IRIDA Observatory, 17A Prof. Asen Zlatarov str., Sofia, Bulgaria\\
                \and
   Institute of Astronomy and National Astronomical Observatory, Bulgarian Academy of Sciences, 72, Tsarigradsko Shose Blvd., 1784 Sofia, Bulgaria\\
   {\small Received ; accepted }}

\abstract{Photometric observations in Sloan \emph{g'} and
\emph{i'} bands of four W UMa stars, NSVS 2244206, NSVS 908513,
CSS J004004.7+385531, VSX J062624.4+570907, are presented. The
light curve solutions reveal that all targets have overcontact
configurations with fillout factor within 0.15--0.26. Their
components are of G-K spectral type and almost in thermal
contacts. They are relatively close also in size and luminosity:
the radius ratios $r_2/r_1$ are within 0.75--0.90; the luminosity
ratios $l_2/l_1$ are within 0.53--0.63. The results of the light
curve solution of CSS~J004004.7+385531 imply weak limb-darkening
effect of its primary component and possible presence of
additional absorbing feature in the system.  }

   \authorrunning{D. Kjurkchieva, V. Popov, D. Vasileva, N. Petrov}            
   \titlerunning{Light curve solutions of four W UMa stars}  
   \maketitle

%
\section{Introduction}           
\label{sect:intro}

W UMa-type binaries consist of two cool stars (F, G, K spectral
type) in contact with each other, surrounded by a common
convective envelope lying between the inner and outer critical
Roche surfaces. In result their components possess almost
identical surface brightness, i.e. temperature
(\citealt{Lucy+1968}, \citealt{Lucy+1976}).

\newpage
The periods of W UMa binaries are in the range 0.22--0.70 days.
They present numerous family: around of 1/500--1/130 MS stars in
the solar neighborhood (\citealt{Rucinski+2002}). There are many
studies on them (\citealt{Liu+etal+2011},
\citealt{Qian+etal+2013}, \citealt{Liao+etal+2014}, etc.) but a
complete theory of their origin, structure, evolution and future
fate still lacks.

The most present theoretical models explain the formation of
(short-period) contact systems by the systematic angular momentum
loss (AML) in initially detached binaries with orbital periods of
a couple of days, due to the magnetized stellar winds and tidal
coupling (\citealt{Vilhu+1981}, \citealt{Rahunen+1982},
\citealt{Stepien+1995}). But according to
\cite{Pribulla+Rucinski+2006} a third (distant) companion is
necessary for formation of systems with a period under 1 day.

There are two models of the evolution during the contact phase
itself. The thermal relaxation oscillation (TRO) model assumes
that each component of the binary is out of thermal equilibrium
and its size oscillates around the inner Roche lobe
(\citealt{Lucy+1976}, \citealt{Flannery+1976},
\citealt{Webbink+1977}, \citealt{Yakut+Eggleton+2005}). The binary
spends a part of its present life in contact (when both stars fill
their Roche lobes and mass flows from the secondary to the
primary) and the rest as a semi-detached binary (when only the
primary fills its Roche lobe and mass flows from the primary to
the secondary), slowly evolving towards an extreme mass ratio
system. TRO model explains well the geometry of the W UMa-type
stars: the primary component is an ordinary MS star and the
secondary is also a MS star but swollen to its Roche lobe by
energy transfer. The main problem of the TRO model is the
mechanism of the energy transfer. The alternative model
(\citealt{Stepien+2004}, \citealt{Stepien+2006},
\citealt{Stepien+2009}, \citealt{Stepien+2011}) assumes that mass
transfer occurs with the mass ratio reversal, similarly as in
Algol-type binaries, following the Roche lobe overflow (RLOF) by
the massive component. The contact configuration is formed
immediately after that or after some additional AML. Each
component is in thermal equilibrium and the large size of the
currently less massive component results from its advanced
evolutionary stage (its core is hydrogen depleted).

The final products of the W UMa-type evolution are also debatable
(\citealt{Li+etal+2007}, \citealt{Eker+etal+2008}). It is supposed
that they may become: single blue stragglers (by merging of the W
UMa components as a result of high rate of angular momentum loss);
two brown dwarfs (\citealt{Li+etal+2007}) or two stars with very
low mass (if mass-loss rate is very high).

The W-phenomenon is another unresolved problem and interesting
peculiarity of the W UMa stars appearing by the lower apparent
surface brightness of the more massive components of the W-type
systems (\citealt{Binnendijk+1970}). They are recognized by the
primary minima which are occultations (indicating that the small
components are the hotter ones). It was suspected that this effect
is due to a large coverage of the primary with cool, dark spots
reducing significantly its apparent luminosity
(\citealt{Eaton+etal+1980}, \citealt{Stepien+1980},
\citealt{Hendry+etal+1992}), but this explanation was not entirely
convincing.
\cite{Gazeas+Niarchos+2006} suggested that subtype A systems have
higher total angular momentum (AM) and can evolve into subtype W
which is the opposite of the earlier conclusion.

\newpage
Besides their key role for understanding of the stellar evolution,
the contact binaries are natural laboratories to study important
astrophysical processes: interaction of stellar winds; magnetic
activity; mass, energy and angular momentum transfer and loss;
phenomenon ''mass ratio reversal''; merging or fusion of the stars
(\citealt{Martin+etal+2011}). The period-color-luminosity relation
of the contact binary stars are an useful tool for distance
determination (\citealt{Rucinski+1994}, \citealt{Rucinski+1996};
\citealt{Rucinski+Duerbeck+1997};
\citealt{Klagyivik+Csizmadia+2004}; \citealt{Eker+etal+2008}).

Hence, the study of the properties of the W UMa stars and their
variety is important for the modern astrophysics. But the
statistics of the most interesting W UMa stars, those with short
periods, is still quite poor (\citealt{Terrell+etal+2012}) mainly
due to their faintness (they are late stars).


In this paper we present photometric observations and light curve
solutions of four short-period W UMa stars: NSVS 2244206, NSVS 908513, CSS
J004004.7+385531 $\equiv$ 2MASS J00400476+3855318 $\equiv$ GSC
02797-00705 $\equiv$ UCAC4-645-002474; VSX J062624.4+570907
$\equiv$ 2MASS J06262444+5709075 $\equiv$ CSS J062624.5+570907
$\equiv$ GSC 03772-01134. Table~1 presents their coordinates and
available (preliminary) information for their light variability.


\begin{table*}[tp]
\begin{center}
\caption[]{Previous information for our targets \label{Tab1}}
\footnotesize
 \begin{tabular}{ccccccccc}
  \hline\hline
  \noalign{\smallskip}
Name         & RA          & Dec         & Period & Epoch    & V    & Ampl & Type &  Ref  \\
             &             &             & [d]    &  [d]     & [mag]& [mag]&      & \\
  \hline
  \noalign{\smallskip}
NSVS 2244206         & 06 06 20.21 & +65 07 21.0 & 0.280727 & -          & 11.997& 0.32 & EB/EW  & 1 \\
NSVS 908513          & 12 30 39.36 & +83 23 07.8 & 0.399592 & -          & 11.772& 0.53 & EB/EW  & 1  \\
CSS J004004.7+385531 & 00 40 04.73 & +38 55 31.9 & 0.251206 & -          & 13.95 & 0.72 & EW     & 2  \\
VSX J004004.4+385513 & 00 40 04.40 & +38 55 13.6 & 0.251206 & 2451359.756& 13.42 & 0.34 & EW     & 3  \\
VSX J062624.4+570907 & 06 26 24.43 & +57 09 07.4 & 0.280628 & 2455162.795& 12.72 & 0.68 & EW     & 3   \\
\hline\hline
 \end{tabular}
\end{center}
References: 1 -- \cite{Gettel+etal+2006}; 2 -- \cite{Drake+etal+2014}; 3 --
\cite{Wozniak+etal+2004};
\end{table*}

\begin{table*}[tp]\footnotesize
\begin{center}
\caption[]{Journal of the Rozhen photometric observations
\label{t2}}
 \begin{tabular}{ccccc}
\hline\hline
Target&  Date        & Exposure ($g',i'$) & Number ($g',i'$) & Error ($g',i'$) \\
      &              & [sec]              &                  & [mag]              \\
  \hline
NSVS 2244206         & 2015 Jan 8  & 60, 90 & 64, 64   & 0.004, 0.004\\
                     & 2015 Jan 11 & 60, 90 & 78, 78   & 0.003, 0.004\\
                     & 2015 Jan 12 & 60, 90 & 91, 84   & 0.004, 0.004\\
                     & 2015 Jan 15 & 60, 90 & 147, 171 & 0.003, 0.003\\
 \hline
NSVS 908513          & 2015 Mar 31 & 60, 90 & 55, 55   & 0.003, 0.004\\
                     & 2015 Apr 11 & 60, 90 & 106, 144 & 0.002, 0.003\\
                     & 2015 Apr 16 & 60, 90 & 138, 137 & 0.004, 0.005\\
 \hline
CSS J004004.7+385531 & 2014 Nov 10 & 120, 120 & 110, 112 & 0.014, 0.014\\
                     & 2014 Nov 20 & 120, 120 & 16, 16 & 0.011, 0.015\\
                     & 2014 Nov 22 & 120, 120 & 44, 60 & 0.012, 0.014\\
                     & 2014 Nov 26 & 120, 120 & 74, 66 & 0.008, 0.010\\
 \hline
VSX J062624.4+570907 & 2014 Dec 24 & 150, 150 & 118, 117   & 0.003, 0.006\\
   \hline\hline
\end{tabular}
\end{center}
\end{table*}

\begin{table*}[tp]
\begin{center}
\caption[]{Coordinates and magnitudes of the standard and check stars \label{t3}}
\footnotesize
 \begin{tabular}{ccccccccc}
   \hline
  \noalign{\smallskip}
Label  &        Star ID     & Other designations&   RA          &  Dec              &  \emph{g'} &  \emph{ i' }\\
  \hline\noalign{\smallskip}
Target 1& NSVS 2244206 & UCAC4 776-023507 & 06 06 20.21 & +65 07 21.0 & 11.977 & 11.144 \\
Chk & UCAC4 776-023451 & GSC 04103-01161 & 06 05 04.84 &  +65 07 17.21 & 11.736 & 11.367 \\
C1 & UCAC4 777-021920 & GSC 04103-00028 & 06 06 58.03 &  +65 18 15.55 & 13.905 & 12.452 \\
C2 & UCAC4 777-021910 & GSC 04103-00908 & 06 06 39.95 &  +65 17 56.15 & 13.270 & 12.842 \\
C3 & UCAC4 777-021896 & GSC 04103-00086 & 06 06 31.01 &  +65 15 33.41 & 12.338 & 11.895 \\
C4 & UCAC4 776-023478 & GSC 04103-01089 & 06 05 43.92 &  +65 08 32.90 & 13.219 & 12.628 \\
C5 & UCAC4 776-023468 & GSC 04103-01386 & 06 05 27.41 &  +65 03 00.31 & 12.335 & 11.802 \\
C6 & UCAC4 776-023541 & GSC 04103-00274 & 06 06 57.18 &  +65 00 11.24 & 13.121 & 12.600 \\
C7 & UCAC4 775-024696 & GSC 04103-00798 & 06 07 27.43 &  +64 59 18.52 & 12.602 & 11.433 \\
C8 & UCAC4 776-023570 & GSC 04103-00108 & 06 07 39.47 &  +65 04 53.91 & 13.765 & 13.233 \\
Target 2 & NSVS 908513 & UCAC4 867-006050 & 12 30 39.36 & +83 23 07.8 & 11.772 & 11.105 \\
Chk & UCAC4 867-006038 & GSC 04633-01779 & 12 29 15.53 &  +83 20 58.94 & 13.285 & 12.466 \\
C1 & UCAC4 867-006016 & GSC 04633-01264 & 12 25 15.86 &  +83 16 24.90 & 11.335 & 10.733 \\
C2 & UCAC4 867-006026 & GSC 04633-01365 & 12 26 44.73 &  +83 15 13.74 & 11.088 & 10.747 \\
C3 & UCAC4 867-006048 & GSC 04633-01496 & 12 30 02.21 &  +83 15 12.10 & 13.035 & 11.580 \\
C4 & UCAC4 867-006056 & GSC 04633-01284 & 12 31 17.68 &  +83 22 11.16 & 13.786 & 12.918 \\
C5 & UCAC4 867-006057 & GSC 04633-01516 & 12 31 39.72 &  +83 21 50.46 & 11.334 & 10.880 \\
C6 & UCAC4 868-005899 & GSC 04633-01339 & 12 24 15.66 &  +83 27 36.10 & 12.305 & 11.610 \\
C7 & UCAC4 868-005922 & GSC 04633-01435 & 12 28 54.40 &  +83 26 09.13 & 12.605 & 11.971 \\
C8 & UCAC4 868-005969 & GSC 04633-01360 & 12 35 27.00 &  +83 26 58.19 & 12.049 & 10.779 \\
C9 & UCAC4 868-005942 & GSC 04633-01599 & 12 31 56.99 &  +83 29 43.30 & 12.548 & 11.966 \\
C10 & UCAC4 868-005965 & GSC 04633-01703 & 12 34 40.00 &  +83 32 27.62 & 12.985 & 12.260 \\
C11 & UCAC4 868-005971 & GSC 04633-01424 & 12 35 37.57 &  +83 33 21.72 & 13.667 & 12.479 \\
C12 & UCAC4 868-005980 & GSC 04633-01616 & 12 37 02.02 &  +83 32 47.47 & 11.259 & 10.850 \\
Target 3 & CSS J004004.7+385531 & UCAC4 645-002474 & 00 40 04.73 & +38 55 31.9 & 14.531 & 13.314 \\
Chk & UCAC4 645-002460 & GSC 02784-01714 & 00 39 51.35 &  +38 55 14.15 & 14.129 & 13.642 \\
C1 & UCAC4 645-002459 & GSC 02784-00660 & 00 39 51.08 &  +38 52 47.94 & 13.984 & 13.560 \\
C2 & UCAC4 645-002487 & GSC 02797-00707 & 00 40 16.35 &  +38 51 36.13 & 14.213 & 13.208 \\
C3 & UCAC4 645-002499 & GSC 02797-00841 & 00 40 26.60 &  +38 55 17.97 & 14.117 & 13.528 \\
C4 & UCAC4 645-002488 & GSC 02797-00759 & 00 40 16.49 &  +38 57 06.96 & 14.186 & 13.674 \\
C5 & UCAC4 645-002434 & GSC 02784-00090 & 00 39 22.45 &  +38 55 07.13 & 13.935 & 13.808 \\
C6 & UCAC4 645-002509 & GSC 02797-00433 & 00 40 41.68 &  +38 55 19.64 & 14.343 & 13.750 \\
C7 & UCAC4 644-002468 & GSC 02784-01325 & 00 39 34.96 &  +38 42 54.15 & 13.725 & 13.214 \\
C8 & UCAC4 644-002496 & GSC 02784-01757 & 00 39 59.82 &  +38 43 47.26 & 14.275 & 13.392 \\
Target 4 & VSX J062624.4+570907 & UCAC4 736-046853 & 06 26 24.43 & +57 09 07.4 & 13.413 & 12.469 \\
Chk & UCAC4 737-044764 & GSC 03773-00028 & 06 27 28.22 &  +57 14 51.82 & 13.912 & 12.355 \\
C1 & UCAC4 737-044774 & GSC 03773-00266 & 06 27 41.41 &  +57 16 07.39 & 14.768 & 14.029 \\
C2 & UCAC4 737-044766 & GSC 03773-00279 & 06 27 30.58 &  +57 16 07.12 & 14.517 & 13.872 \\
C3 & UCAC4 737-044778 & GSC 03773-00030 & 06 27 48.83 &  +57 18 39.37 & 13.565 & 13.001 \\
C4 & UCAC4 736-046930 & GSC 03773-00302 & 06 27 53.71 &  +57 11 14.10 & 13.344 & 12.349 \\
C5 & UCAC4 736-046874 & GSC 03773-00004 & 06 26 46.16 &  +57 11 30.66 & 13.964 & 13.318 \\
C6 & UCAC4 736-046859 & GSC 03772-01339 & 06 26 33.14 &  +57 04 49.70 & 13.514 & 12.656 \\
C7 & UCAC4 736-046863 & GSC 03772-00447 & 06 26 36.42 &  +57 02 51.35 & 14.111 & 13.185 \\
C8 & UCAC4 736-046818 & GSC 03772-01556 & 06 25 48.21 &  +57 08 54.91 & 13.787 & 12.977 \\
\hline\hline
 \end{tabular}
\end{center}
\end{table*}

\section{Observations}

Our CCD photometric observations of the targets in Sloan \emph{g',
i'} bands were carried out at Rozhen Observatory with the 30-cm
Ritchey Chretien Astrograph (located into the \emph{IRIDA South}
dome) using CCD camera ATIK 4000M (2048 $\times$ 2048 pixels, 7.4
$\mu$m/pixel, field of view 35 x 35 arcmin). Information for our
observations is presented in Table~2.

The photometric data were reduced by {\textsc{AIP4WIN2.0}
(\citealt{Berry+Burnell+2005}). We performed aperture ensemble
photometry with the software \textsc{VPHOT} using more than six
standard stars in the observed field of each target. The
coordinates and magnitudes of the standard and check stars (Table
3) were taken from the catalogue UCAC4
(\citealt{Zacharias+etal+2012}).

We established that there are two close objects,
CSS~J004004.7+385531 and VSX~J004004.4+385513, with the same
periods and types of variability in the VSX database (Table 1).
Our observations revealed that the true variable is CSS
J004004.7+385531 while VSX~J004004.4+385513 is a stationary star
(Fig. 1).

\begin{figure}
   \centering
   \includegraphics[width=8 cm, angle=0]{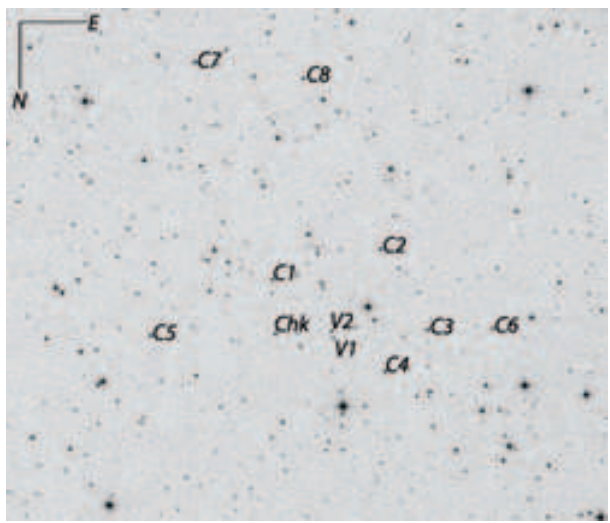}
   \caption{The field of CSS J004004.7+385531(V1) with the close
   star VSX J004004.4+385513 (V2) wrongly considered as a variable}
   \label{Fig1}
   \end{figure}

\begin{figure}
   \centering
   \includegraphics[width=0.55\columnwidth]{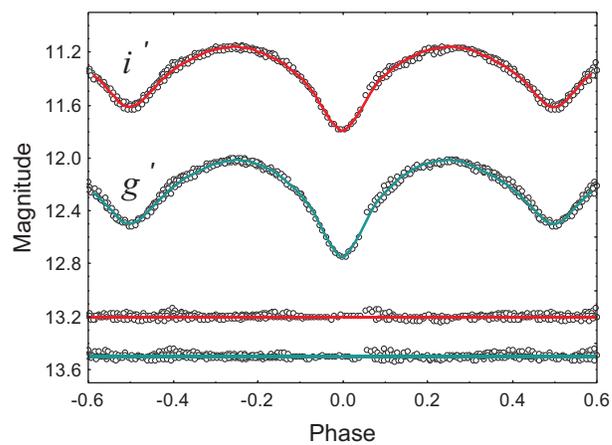}
   \caption{Top: the folded light curves of NSVS 2244206 and their fits; Bottom: the corresponding residuals
   (shifted vertically by different number to save space).
   Color version of this figure is available in the online journal.}
   \label{Fig2}
   \end{figure}

\begin{figure}
   \centering
   \includegraphics[width=0.55\columnwidth]{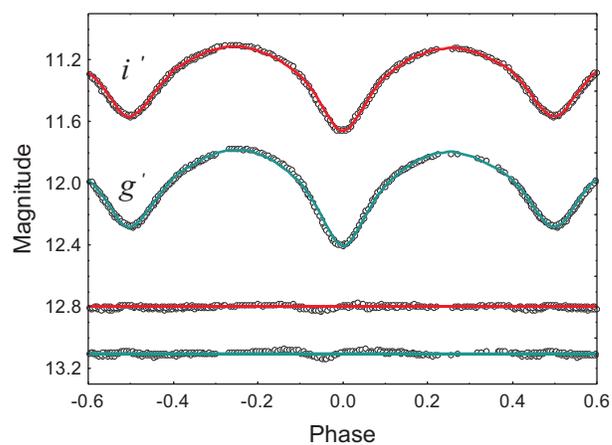}
   \caption{Same as Fig. 2 for NSVS 908513}
   \label{Fig3}
   \end{figure}

\begin{figure}
   \centering
   \includegraphics[width=0.55\columnwidth]{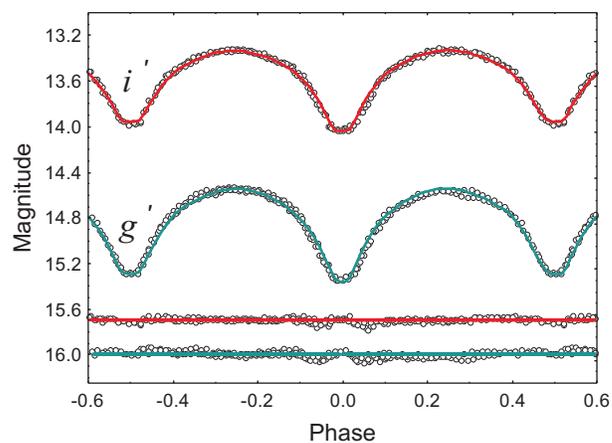}
   \caption{Same as Fig. 2 for CSS J004004.7+385531}
   \label{Fig4}
   \end{figure}

\begin{figure}
   \centering
   \includegraphics[width=0.55\columnwidth]{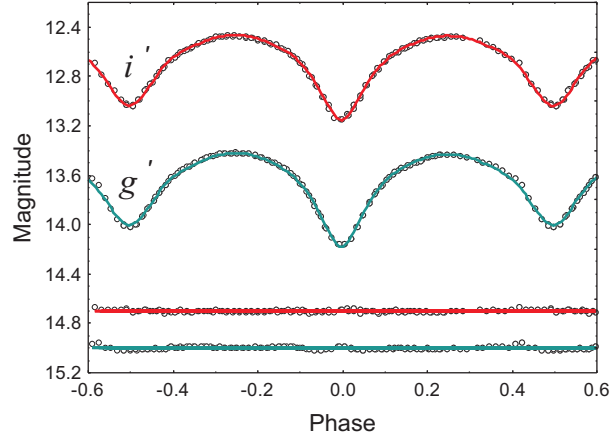}
   \caption{Same as Fig. 2 for VSX J062624.4+570907}
   \label{Fig5}
   \end{figure}

We determined the times of the individual minima (Table 4)
by the method of \cite{Kwee+Woerden+1956}.

\begin{table}
\begin{center}
\caption[]{Times of minima of our targets \label{Tab1}}
\footnotesize
\begin{tabular}{ccc}
\hline\hline
\noalign{\smallskip}
Target & MinI & MinII \\
\hline
\noalign{\smallskip}
NSVS 2244206 & - & 2457033.51298\\
& - & 2457034.35629 \\
& - & 2457038.28497 \\
& 2457038.42614 & - \\
\hline
NSVS 908513 & - & 2457113.31730 \\
& 2457124.30454 & 2457124.50608 \\
& 2457129.50051 & 2457129.30045 \\
\hline
CSS J004004.7+385531 & 2456972.34036 & 2456972.46714 \\
& 2456972.59113 & 2456982.26682 \\
& 2456984.39854 & 2456988.29364 \\
& 2456988.41840 & - \\
\hline
VSX J062624.4+570907 & 2457016.33657 & 2457016.47792 \\
& 2457016.61853 & - \\
\hline\hline
\end{tabular}
\end{center}
\end{table}

\section{Light curve solutions}

We carried out modeling of the photometric data by the code
\textsc{PHOEBE} (\citealt{Prsa+Zwitter+2005}). It is based on the
Wilson--Devinney (WD) code (\citealt{Wilson+Devinney+1971},
\citealt{Wilson+1979}). \textsc{PHOEBE} incorporates all the
functionality of the WD code but also provides a graphical user
interface alongside other improvements, including updated filters
as Sloan ones used in our observations. We apply the traditional
convention the MinI (phase 0.0) to be the deeper light minimum and
the star that is eclipsed at MinI to be the primary (hotter)
component.

\newpage
We determined in advance the mean temperatures $T_{m}$ of the
binaries (Table~5) by their infrared color indices \emph{(J-K)}
from the 2MASS catalog and the calibration color-temperature of
\cite{Tokunaga+2000}. In fact, the determination of stellar
temperatures from the infrared flux is a method first developed by
\cite {Blackwell+Shallis+1977}.

Our procedure of the light curve solutions was carried out in
several stages.

At the first stage we fixed $T_{1}^0$ = $T_{m}$ and searched for fit
varying the secondary temperature $T_{2}$, orbital inclination
$i$, mass ratio $q=m_2/m_1$ and potentials $\Omega_{1,2}$ (and
thus relative radii $r_{1,2}$ and fillout factor \emph{f}). The
fit quality was estimated by the value of $\chi^2$.

\newpage
Coefficients of gravity brightening and reflection effect
appropriate for stars with convective envelopes were adopted.
Initially we used linear limb-darkening law with limb-darkening
coefficients corresponding to the stellar temperatures and
Sloan photometric system (\citealt{Claret+Bloemen+2011}).

In order to reproduce the light curve distortions of the targets
we added cool spots on the stellar surfaces and varied spot
parameters: longitude $\lambda$, latitude $\beta$, angular size
$\alpha$ and temperature factor $\kappa=T_{sp}/T_{st}$.

As a result of the first stage of the light curve solution
we obtained the values $T_{2}^{0}$, $i^{0}$, $\Omega_{1,2}^{0}$, $q$
as well as the spot parameters for each target.

After reaching the best fit we adjusted $T_{1}$ and $T_{2}$ around
the value $T_m$ by the formulae (\citealt{Kjurkchieva+etal+2015})

\begin{equation}
T_1^c=T_{\rm {m}} + \frac{c \Delta T}{c+1}
\end{equation}
\begin{equation}
T_2^c=T_1^c -\Delta T
\end{equation}
where $c=l_2/l_1$ and $\Delta T=T_{1}^{0}-T_{2}^{0}$ are determined from the \textsc{PHOEBE} solution.

Finally, we varied slightly $T_{1}$, $T_{2}$, $i$ and
$\Omega_{1,2}$ around their values $T_{1}^{c}$, $T_{2}^{c}$, $i^{0}$ and
$\Omega_{1,2}^{0}$ and obtained the final \textsc{PHOEBE}
solution.

The first part of Table~5 contains the parameters of our light
curve solutions: mass ratio $q$; orbital inclination $i$;
potentials $\Omega_{1,2}$; fillout factor \emph{f}; stellar
temperatures $T_{1, 2}$; relative radii $r_{1, 2}$; ratio of
relative luminosities $l_{2}/l_{1}$. The errors of these
parameters are the formal \textsc{PHOEBE} errors. Table~6 gives
the obtained spot parameters. The synthetic curves corresponding
to our light curve solutions are shown in Figs. 2-5.

\begin{figure}
\begin{center}
\includegraphics[width=3cm]{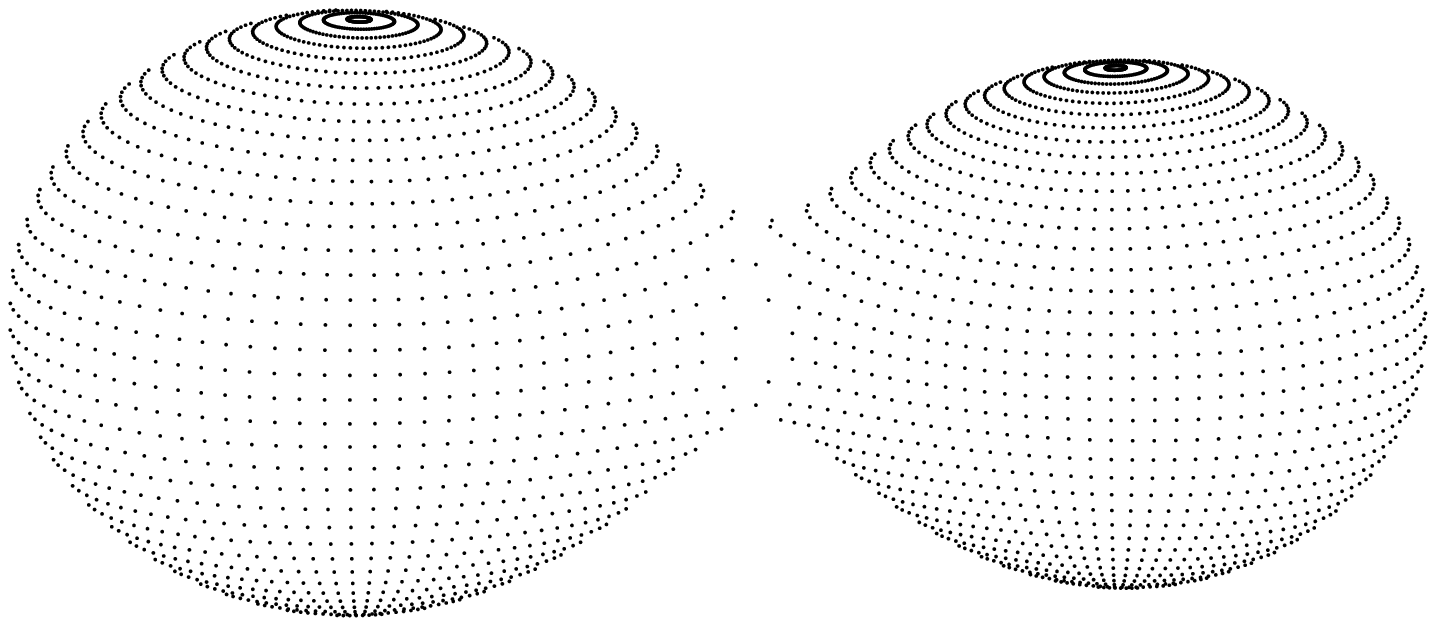}
\includegraphics[width=3cm]{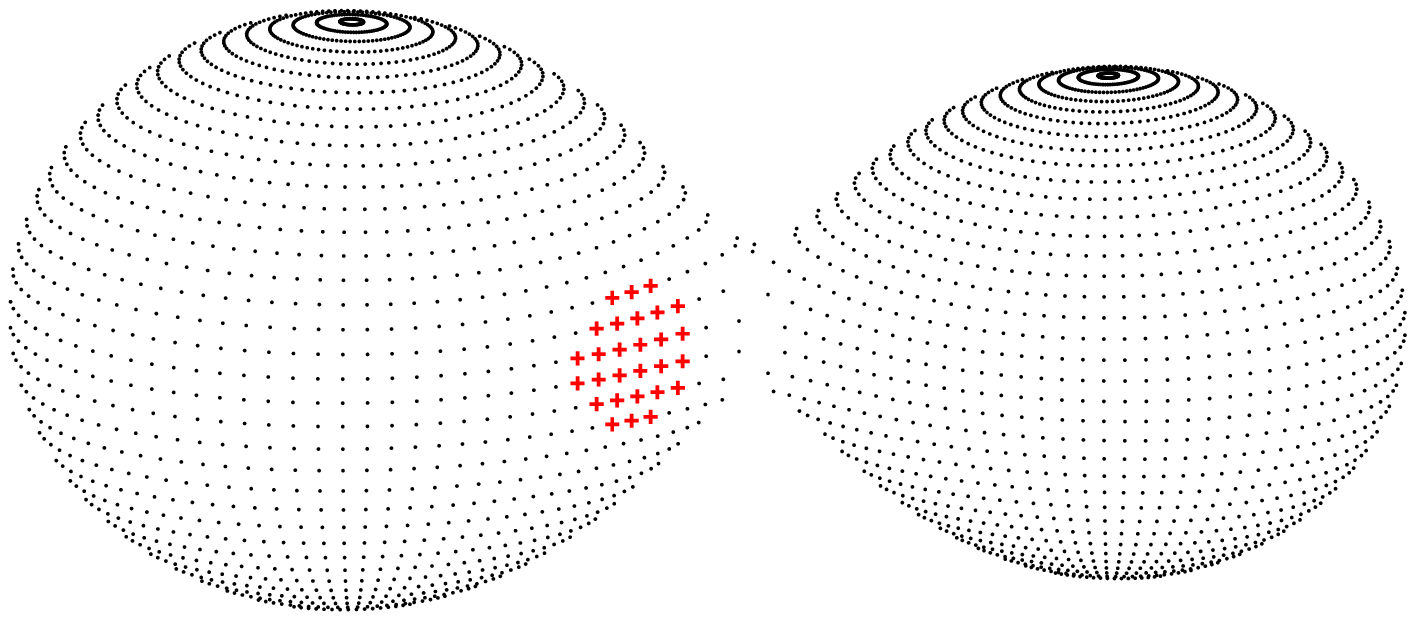}
\includegraphics[width=3cm]{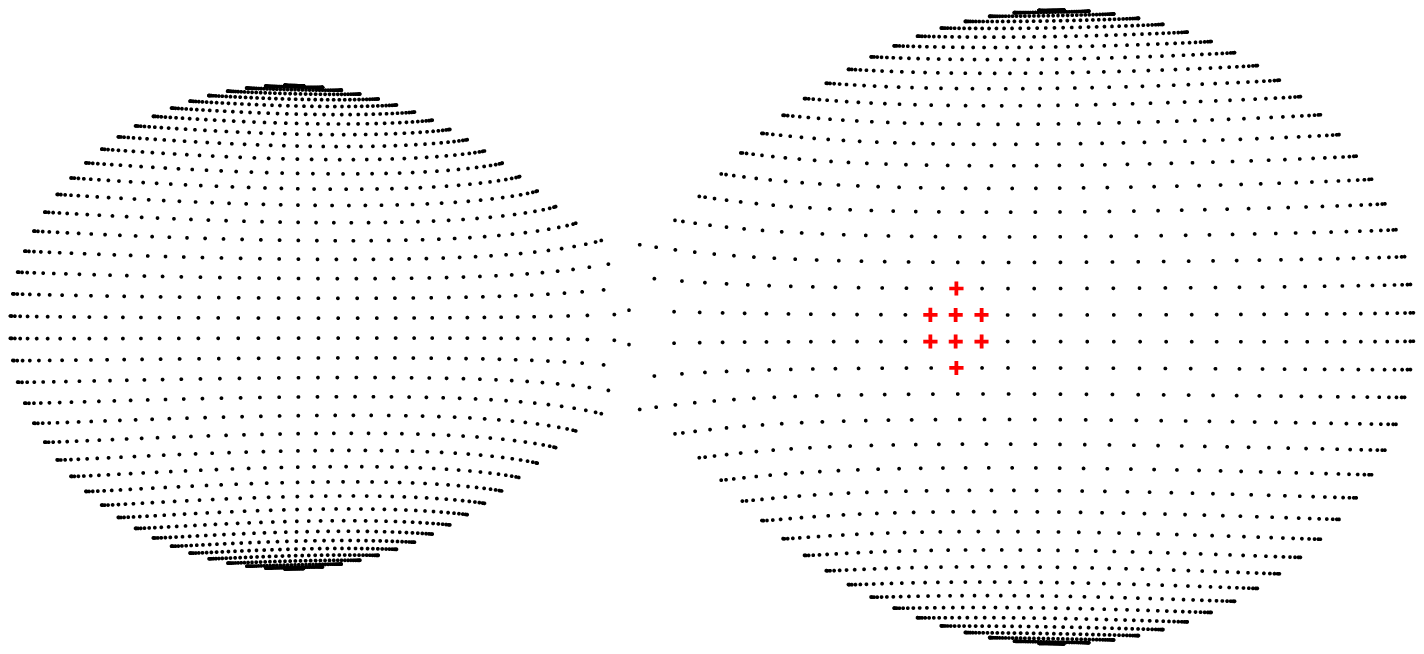}
\includegraphics[width=3cm]{MS2618fig6b.eps}
\caption{From the left to the right 3D configurations of: NSVS 2244206, NSVS 908513, CSS J004004.7+385531, VSX J062624.4+570907}\label{fig4}
\end{center}
\end{figure}

Due to the lack of radial velocity measurements we had not a
possibility to determine reliable values of the global parameters
of the target components. We were able to obtain some estimations
of these quantities by the following procedure.

The primary luminosity $L_1$ was determined by the empirical
relation luminosity-temperature for MS stars. The secondary
luminosity $L_2$ was calculated by the relation $L_2=c L_1$ where
$c=l_2/l_1$ is the luminosity ratio from our light curve solution.

We obtained the orbital separation $a$ (in solar radii) from the
equation
\begin{equation}
\log a = 0.5 \log L_i - \log r_{i} - 2 \log T_{i} + 2 \log
T_{\odot}.
\end{equation}
and then calculated the absolute stellar radii by $R_{i}=ar_{i}$.

\newpage
The total target mass $M$ (in solar units) was calculated from the
third Kepler law
\begin{equation}
M=\frac{0.0134 a^3}{P^2}
\end{equation}
where \emph{P} is in days while \emph{a} is in solar radii. Then
the individual masses $M_i$ were determined by the formulae $ M_1
=  M /(1+q)$ and $ M_2= M- M_1$.

The global parameters of the target components obtained by the
foregoing procedure are given in the second part of Table~5. Their
errors are calculated from the corresponding formulae by the
errors of the quantities of the light curve solutions or
observations.

\begin{table*}[tp]\footnotesize
\begin{center}
\caption{Parameters of the best light curve solutions \label{t4}
(top) and global parameters (bottom) of the targets}
\begin{tabular}{ccccc}
\hline\hline
Star name               &   NSVS 2244206    &   NSVS 908513             &   CSS J004004.7+385531    &   VSX J062624.4+570907            \\
\hline
\emph{q}                &   0.735$\pm$0.003 &   0.709   $\pm$   0.002   &   0.548   $\pm$   0.004   &   0.777   $\pm$   0.002   \\
\emph{i}, ($^{\circ}$)  &   76.42$\pm$0.07  &   75.15   $\pm$   0.03    &   89.77   $\pm$   0.01    &   78.88   $\pm$   0.11 \\
$\Omega_1=\Omega_2$     &   3.1961$\pm$.004 &   3.2     $\pm$   0.002   &   2.9     $\pm$   0.009   &   3.307   $\pm$   0.004   \\
\emph{f}                &   0.260$\pm$0.002 &   0.146   $\pm$   0.004   &   0.206   $\pm$  0.009    &   0.162   $\pm$   0.004   \\
$T_m$, (K)              &   5000            &   5810                    &   4560                    &   5230           \\
$T_1$, (K)              &   5157 $\pm$  36  &   5923    $\pm$   25      &   4560    $\pm$   37      &   5350    $\pm$   20       \\
$T_2$, (K)              &   4702 $\pm$  32  &   5615    $\pm$   23      &   4560    $\pm$   38      &   5044    $\pm$   7         \\
$r_1$                   &   0.429$\pm$0.001 &   0.422   $\pm$   0.001   &   0.449   $\pm$   0.003   &   0.416   $\pm$   0.002   \\
$r_2$                   &   0.376$\pm$0.001 &   0.363   $\pm$   0.001   &   0.344   $\pm$   0.004   &   0.372   $\pm$   0.002   \\
$l_2/l_1$               &   0.5301          &   0.5946                  &   0.5883                  &   0.6314                  \\
\hline
$L_{1}^{bol}$           &   0.519$\pm$0.023 &   1.031   $\pm$   0.023   &   0.254   $\pm$   0.04    &   0.604   $\pm$   0.073   \\
$L_{2}^{bol}$           &   0.338$\pm$0.045 &   0.646   $\pm$   0.046   &   0.149   $\pm$   0.062   &   0.421   $\pm$   0.096   \\
\emph{a}, (R$_{\odot}$) &   2.214$\pm$0.06  &   2.312   $\pm$   0.066   &   1.798   $\pm$   0.055   &   2.23    $\pm$   0.067   \\
$R_1$, (R$_{\odot}$)    &   0.951$\pm$0.028 &   0.976   $\pm$   0.03    &   0.808   $\pm$   0.03    &   0.927   $\pm$   0.032   \\
$R_2$, (R$_{\odot}$)    &   0.833$\pm$0.032 &   0.839   $\pm$   0.033   &   0.619   $\pm$   0.026   &   0.829   $\pm$   0.027   \\
$M_1$, (M$_{\odot}$)    &   1.064$\pm$0.084 &   0.607   $\pm$   0.051   &   0.798   $\pm$   0.07    &   1.061   $\pm$   0.094   \\
$M_2$, (M$_{\odot}$)    &   0.782$\pm$0.067 &   0.430   $\pm$   0.038   &   0.437   $\pm$   0.043   &   0.825   $\pm$   0.077   \\
\hline\hline
\end{tabular}
\end{center}
\end{table*}

\section{Analysis of the results}

The analysis of the light curve solutions of our short-period W
UMa stars led to several important results.

(1) CSS J004004.7+385531 reveals total eclipses while the remained
three targets undergo partial eclipses.

(2) The temperatures of the stellar components of the targets
correspond to G-K spectral type (Table~5). The temperature
differences of their components do not exceed 450 K while the
components of CSS J004004.7+385531 are in precise thermal contact.

(3) The targets have overcontact configurations which fillout
factors \emph{f} are in the range 0.15--0.26 (Table 5). It should
be pointed out that the preliminary classification of NSVS 2244206
and NSVS 908513 was EB/EW (Table 1) but our observations and light
curve solutions led to the conclusion that their configurations
are overcontact (Fig. 6).

\newpage
(4) The target components are relatively close in size and
luminosity: the size ratios $r_2/r_1$ are within 0.75--0.90; the
luminosity ratios $l_2/l_1$ are within 0.53--0.63.

(5) We observed slightly different levels of the two
quadratures (O'Connell effect) of three our targets. They were
reproduced by small cool spots (Table 6) on their primaries. We
obtained solutions with the same fit quality for combinations of
slightly different spot sizes (within 1$^{o}$) and latitudes ($\pm
25^{o}$ around the stellar equator). Table~6 presents the
parameters of the equatorial spots whose angular sizes have
minimum values.

(6) The residuals of CSS J004004.7+385531 are bigger than those of
the other three targets that is expected taking into account that
this totally-eclipsed binary is the faintest member of our sample
(Table 1). But the residuals are biggest at its primary eclipse
because the synthetic eclipse is narrower than the observed one
(Fig. 4). The reason for this discrepancy is that the observed
primary eclipse of CSS J004004.7+385531 turns out slightly wider
than the secondary one. This may due to some additional structure
in the system which presence cannot be taken into account from the
software for light curve synthesis: equatorial bulge around the
less massive component (accretor) formed as a result of the
transferred mass; disk-like feature; clouds at some Lagrangian
points (result of previous nonconservative mass transfer,
\citealt{Stepien+Kiraga+2013}).

(7) We managed to reproduce the almost flat bottom of the
primary eclipse of CSS J004004.7+385531, especially in \emph{i'}
band, only by very small limb-darkening coefficient of 0.18, a
value considerably smaller than that corresponding to its
temperature. This formally means faint limb-darkening effect, i.e.
almost homogeneous stellar disk of the primary component. There
are two possible reasons for this effect. The first one is that
the theory and corresponding codes for light curve synthesis
cannot take into account precisely the limb-darkening effect for
overcontact binaries with strongly distorted components whose
photospheres are deeply inside the envelopes. The second reason
for the flat bottom of the primary eclipse might be light
contribution of optically thick region around L1 that is covered
by the secondary component at the primary eclipse. Such an
argument refers just for CSS J004004.7+385531 that undergoes
almost central eclipse.

(8) Quite often photometric solutions of W UMa light curves
appear to be ambiguous since both A and W configurations can fit well
the observations (\citealt{vanHamme+1982a}; \citealt{Lapasset+Claria+1986}).
The mass ratio of the W UMa binaries is important
parameter for their W/A subclassification. But the rapid rotation
of their components does not allow to obtain precise spectral mass
ratio from measurement of their highly broadened and blended
spectral lines (\citealt{Bilir+etal+2005};
\citealt{Dall+Schmidtobreick+2005}). As a result the W/A
subclassification of the W UMa binaries
is made mainly on the widely-accepted empirical relation
''spectral type -- mass'' (\citealt{vanHamme+1982a}; \citealt{Lapasset+Claria+1986}):
the G-K binaries are of W subtype while A and earlier F
binaries are of A subtype.


\newpage
Particularly, our targets are faints objects and we have
obtained only their photometric mass ratios \emph{q} by
varying this parameter within the range 0.1-2.5.
Thus, we obtained a pair solutions for each target with
close quality: (a) A-subtype solution with parameters $q^A < 1$,
$r_1^A$, $r_2^A < r_1^A$; (b) W-subtype solution with parameters
$q^W \approx 1/q^A$, $r_1^W \approx r_2^A$, $r_2^W \approx r_1^A$.
To choose one of them we introduced the relative difference
$\Delta Q$ (in percentage) of the solution quality \emph{Q(W)}
corresponding to the W-subtype configuration and the solution
quality \emph{Q(A)} corresponding to the A-subtype configuration
\begin{equation}
\Delta Q = \frac{Q(W)-Q(A)}{Q(A)}   .
\end{equation}
For all our targets we established $\Delta Q \geq 3 \%$,
i.e. the A solutions are better than the W solutions. This allowed
us to assume that A subtype is more probable subclassification of
our targets than the W subtype. Additional considerations for this
choice were: (i) the W solutions did not reproduce so good the
observed eclipse depths; (ii) the big fillout factors of our
targets are inherent to A subtype binaries
(\citealt{vanHamme+1982a}, \citealt{vanHamme+1982b});
(iii) the A solutions are quite sensible to the mass ratio (Fig. 7).

Hence, our targets could be assigned to the exceptions
from the statistical relation ''spectral type -- mass''. The
possible reason for this discrepancy may be that this relation is
derived for W UMa binaries with periods $>$ 0.3 d (\citealt{vanHamme+1982a}, \citealt{vanHamme+1982b}).

\begin{table}\footnotesize
\begin{center}
\caption[]{Parameters of the cool spots on the targets
 \label{t5}}
  \begin{tabular}{ccccc}
\hline\hline
Target          & $\beta$ & $\lambda$ & $\alpha$ & $\kappa$    \\
  \hline
NSVS 908513          & 90 & 330 & 13 & 0.90   \\
CSS J004004.7+385531 & 90 &  70 & 7  & 0.90   \\
VSX J062624.4+570907 & 90 & 270 & 10 & 0.90    \\
\hline\hline
\end{tabular}
\end{center}
\end{table}

\begin{figure}
   \centering
   \includegraphics[width=4.5cm, angle=0]{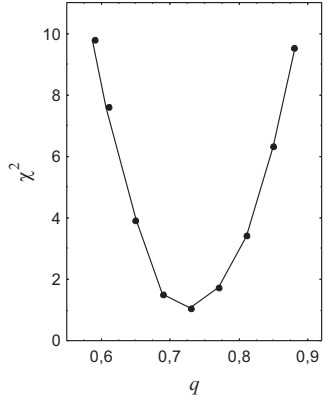}
   \caption{Sensibility of our light curve solution of NSVS 2244206 (measured by $\chi^2$) to the mass ratio
   (the rest parameters last fixed at their final values)}
   \label{Fig7}
   \end{figure}

\newpage
\section{Conclusion}

We obtained light curve solutions of four short-period W UMa
binaries which main results are as follows.

(1) The temperatures of the stellar components of the targets
correspond to G-K spectral type and they are almost in thermal
contacts.

(2) All targets are overcontact configurations with
fillout factor within 0.15--0.26.

(3) The target components are relatively close in size and
luminosity: the size ratios $r_2/r_1$ are within 0.75--0.90; the
luminosity ratios $l_2/l_1$ are within 0.53--0.63.

(4) The results of the light curve solution of
CSS~J004004.7+385531 imply weak limb-darkening effect of its
primary component and possible presence of additional absorbing
feature in the system.

This study adds new four systems with estimated parameters to the
family of short-period binaries. They could help to improve the
statistical relations between the stellar parameters of the
low-massive stars and to better understanding the evolution of
close binaries.

\normalem
\begin{acknowledgements}
The research was supported partly by funds of project RD 08-244 of
Scientific Foundation od Shumen University. It used the SIMBAD
database and NASA Astrophysics Data System Abstract Service. This
research was made possible through the use of the AAVSO
Photometric All-Sky Survey (APASS), funded by the Robert Martin
Ayers Sciences Fund. The authors are grateful to the
anonymous referee for the valuable notes and recommendations.
\end{acknowledgements}

\bibliographystyle{raa}
\bibliography{bibtex}

\end{document}